\documentclass[runningheads]{llncs}
\usepackage{graphicx}
\usepackage{etoolbox}
\makeatletter
\patchcmd{\@verbatim}
  {\verbatim@font}
  {\verbatim@font\small}
  {}{}
\makeatother
  
\begin{document}
\title{A Constraint-based Mathematical\\ Modeling Library in Prolog with\\ Answer Constraint Semantics}
\titlerunning{Prolog Library for Constraint-based Mathematical Modeling}
%
\author{François Fages \orcidID{0000-0001-5650-8266} }
\authorrunning{F. Fages.}
%
\institute{Inria Saclay, Palaiseau, France\\
\email{Francois.Fages@inria.fr}}
\maketitle              
\begin{abstract}
  Constraint logic programming emerged in the late 80's as a highly declarative class of programming languages
  based on first-order logic and theories with decidable constraint languages,
  thereby subsuming Prolog restricted to equality constraints over the Herbrand's term domain.
  This approach has proven extremely successfull in solving combinatorial problems in the industry
  which quickly led to the development of a variety of constraint solving libraries in standard programming languages.
  Later came the design of a purely declarative front-end constraint-based modeling language, MiniZinc,
  independent of the constraint solvers, in order to compare their performances and create model benchmarks.
  Beyond that purpose, the use of a high-level modeling language such as MiniZinc to develop complete applications,
  or to teach constraint programming,
  is limited by the impossibility to program search strategies, or new constraint solvers, in a modeling language,
  as well as by the absence of an integrated development environment for both levels of constraint-based modeling and constraint solving.
  In this paper, we propose to solve those issues by taking Prolog with its constraint solving libraries,
  as a unified relation-based modeling and programming language.
  We present a Prolog library for high-level constraint-based mathematical modeling, inspired by MiniZinc,
  using subscripted variables (arrays) in addition to lists and terms, quantifiers and iterators in addition to recursion,
  together with a patch of constraint libraries in order to allow array functional notations in constraints.
  We show that this approach does not come with a significant computation time overhead,
  and presents several advantages in terms of the possibility of focussing on mathematical modeling, getting answer constraints in addition to ground solutions,
  programming search or constraint solvers if needed, and debugging models within a unique modeling and programming environment.

\end{abstract}
\section{Introduction}

Jean-Baptiste Fourier is very well-known for his numerous contributions to mathematics and physics, but much less known 
for being the father of what is called today Constraint Programming.
In two lectures at the French Academy of Sciences given in 1823 and 1824,
he considered the example of determining the region where a given weight can be placed on a triangular table
with constraints on the maximum forces that can be exerted on each leg,
in order to promote a general problem solving method based on,
first, modeling the problem at hand with inequalities over real numbers,
and second, solving them by applying general purpose simplification rules a.k.a.~Fourier Motzkin's elimination rules today.
Quoting him,
``the advantage of this method consists in that it is sufficient, in all cases, to express the conditions of the question, which is easy, and to then combine these expressions, by means of general rules which are always the same; and we thus form the solution which could only have been reached by a series of very complicated reasonings''.
Furthermore, he mentions that these simplification rules presented for linear inequalities apply as well to non-linear inequalities a.k.a.~interval arithmetic:
``if the conditions are expressed by non-linear inequalities, the question does not change its nature,
and can still be treated by the same principles.''.
In his second lecture, he considers the optimisation problem
and describes a method for moving from vertices to vertices of the feasible region polyhedron,
in order to improve the cost up to its optimal value,
a.k.a~the geometrical interpretation of the Simplex algorithm~\cite{LM92jar}.

In the realm of Linear Programming today, several Algebraic Modeling Languages such as OPL, AMPL, Mosel, etc. have been defined
as input modeling languages with a syntax near to mathematical notations for optimization problems,
using indices, sets, algebraic expressions and data handling variables.
Then external solvers interfaced with these modeling languages can be called to actually solve the problem instances
defined by the model and a dataset.

In the realm of Constraint Programming,
MiniZinc is a similarly high-level constraint-based modeling language to model constraint satisfaction and optimization problems
in a solver-independent way~\cite{NSBBDT07cp,zinc}.
A MiniZinc model is usually transformed in a FlatZinc model in which the high-level constructs have been eliminated
and replaced by a flat constraint satisfaction problem
that can be solved by a variety of constraint solvers parsing FlatZinc syntax.
This is the way for instance SICStus-Prolog is interfaced with FlatZinc as a general purpose constraint solver
to solve problems modeled in MiniZinc~\cite{SICS12SICStus423}.
Because of the importance of global constraints, several of them are defined in MiniZinc, kept in FlatZinc,
and handled by the constraint solver, either directly as global constraints if they are implemented as such,
or by decomposition into basic constraints supported by the constraint solver.

Because of the importance of search in constraint programming,
MiniZinc contains some predefined common search heuristic options
that are kept in FlatZinc for their possible interpretation by the constraint solvers,
in addition to solver specific annotations in the MiniZinc model that are similarly kept in FlatZinc~\cite{NSBBDT07cp}.
Nevertheless, the absence of programming constructs in a modeling language like MiniZinc
limits the search strategies that can be tried to solve a complex problem,
which may finally lead to abandon the high-level modeling approach in favor of a lower level constraint programming language.

Furthermore, when it comes to teaching constraint-based modeling and algorithms for decision making,
there is a need to teach constraint solving algorithms and program them as well.
The recourse to a modeling language like MiniZinc makes it possible to focus on the declarative modeling aspects
but cannot be used to show the implementation of constraint solvers, nor implement new constraints.
Because of its roots in first-order logic, Prolog should be a natural choice to address these needs
as both a programming language in its own right, and a modeling language based on relations a.k.a.~constraints.
Nevertheless the historical focus on list data structure,
and the definition of meta-predicates before the advent of constraint logic programming,
make of Prolog a poor modeling language for complex constraint satisfaction problems,
moreover exhibiting incompatibilities of use between some standard meta-predicates and constraint solving libraries.

In this paper we propose solutions to those issues by presenting a general purpose mathematical modeling library\footnote{The Prolog libraries presented here form a pack named \texttt{modeling}, currently available for SWI-Prolog at \url{https://lifeware.inria.fr/wiki/Main/Software\#modeling}.} in Prolog,
based on bounded quantifiers and arrays for subscripted variables,
which is essentially compatible with MiniZinc models and obviously Prolog programming facilities
for programming search, constraint solvers, and using external interfaces.

The next section contains a motivating example for modeling the N-queens problem
and breaking all the symmetries of that placement problem on a square
(i.e.~with respect to the 8 elements of the dihedral group of isometries of order 4).
The recourse to mathematical notations made possible with our modeling library using subscripted variables and bounded quantifiers rather than lists and recursion,
marks a striking gain in declarativity for engineers and students in engineering schools more acquainted to mathematical notations than programming structures.
We provide some performance figures which show that this is achieved with no computation time overhead.

In Sec.~\ref{zinc} we compare our modeling library in Prolog to the solver-independent
constraint-based modeling language MiniZinc~\cite{NSBBDT07cp,zinc}.
We show the large compatibility between both modeling languages,
and describe the advantages of the Prolog library approach in terms of ability to compute answer constraints, not just ground solutions,
a definite advantage of constraint logic programming~\cite{JL87popl} illustrated there with Fourier's example,
as well as to program search and debug models in a unique modeling and programming environment.

In Sec.~\ref{lib} we describe the main predicates of our modeling library for arrays and iterator meta-predicates
for answer constraint semantics.
The development of such a library raises however a number of subtle issues
that cannot be solved with standard Prolog meta-predicates that have been designed before the advent of constraints.
This leads us to make a proposal for extending the specification of the behaviour of ISO-Prolog meta-predicates
to attributed variables, in the realm of Prolog answer constraint semantics.
Finally, we conclude on the importance of keeping the development of Prolog systems
in both perspectives of a relation-based declarative programming language and a constraint-based mathematical modeling language.

\section{Motivating Example}\label{queens}

\subsection{List Recursion versus Array Iteration in the N-Queens Problem}

The N-queens puzzle consists in placing N queens on an NxN chessboard such that they do not attack each other,
i.e.~no pair of queens is on the same line or diagonal.
This is a classical example used in constraint programming to illustrate
the power of the active use of constraints to prune the search tree when they are posted in advance
in the paradigm of ``constrain and generate'', by opposition to pure backtracking ``generate and test'' programs limited to small size problems.
This makes it possible for instance to place by search 100 queens in 0.25s CPU time on a standard laptop,
a performance out of reach of search methods without active use of constraints to filter the domains of all variables during search.

\begin{figure}
  \centering\includegraphics[width=0.45\textwidth]{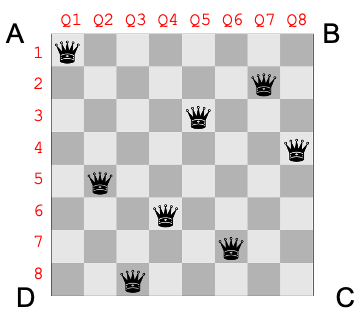}
\caption{Placement of 8 queens on a 8x8 chessboard with no two queens in the same column, row or diagonal, modeled as a constraint satisfaction problem,
  with 8 variables $Q_1,\ldots,Q_8$, representing the queens say in each column,
  taking integer values in the interval $[1,8]$ representing the rows where they are placed.}\label{fig:queens}
\end{figure}

The standard definition of this problem by Prolog clauses with constraints over the integers
uses a list of N variables representing the queens (say per column as in Fig.~\ref{fig:queens})
with a finite domain of possible values in the interval [1,N] (representing their position in the rows)
and posts the no-attack constraints by double recursion on that list of variables,
before enumerating the values in the domain of the variables (predicate \verb|labeling/2|
with \verb|ff| first-fail smallest-domain variable and-choice heuristics),
as follows:

\begin{verbatim}
:- use_module(library(clpfd)).

queens(N, Queens):-
  length(Queens, N),
  Queens ins 1..N,
  safe(Queens),
  labeling([ff], Queens).

safe([]).
safe([QI | Tail]) :-
  noattack(Tail, QI, 1),
  safe(Tail).

noattack([], _, _).
noattack([QJ | Tail], QI, D) :-
  QI #\= QJ, 
  QI #\= QJ + D,
  QI #\= QJ - D,
  D1 #= D + 1,
  noattack(Tail, QI, D1).
\end{verbatim}

The double recursion on the lists of queens and of their successor queens on the right,
is standard in Prolog where the list data structure is promoted, mainly for historical reasons including comparison to Lisp,
and where most built-in predicates are defined on lists.

In algebraic modeling languages however, mathematical notations using subscripted variables,
with iteration over subscripts instead of recursion, are usually preferred for better declarativity.
This is what offers our mathematical modeling library in Prolog.
The double list recursion for posting the constraints between each pair of queens in the program above
can thus be replaced by iteration on the indices of an array of decision variables as follows:

\begin{verbatim}
:- use_module(modeling).

queens(N, Queens):-
  int_array(Queens, [N], 1..N),
  for_all([I in 1..N-1, D in 1..N-I],
          (Queens[I] #\= Queens[I+D],
           Queens[I] #\= Queens[I+D]+D,
           Queens[I] #\= Queens[I+D]-D)),
  satisfy(Queens).
\end{verbatim}

Compared to the previous Prolog program using lists and recursion, 
the advantage in terms of declarativity of the constraint-based model using subscripted variables should be clear,
at least for engineers or students more familiar with mathematical notations than programming structures.

\subsection{Breaking Symmetries in the N-Queens Problem}

The gain in declarativity becomes even more striking however, when it comes to breaking all symmetries of the chessboard square
with respect of the 8 isometries of the dihedral group of order 4 for a square [A,B,C,D] as in Fig.~\ref{fig:queens},
using lexicographic ordering constraints~\cite{Walsh06cp}, namely
\begin{itemize}
\item the reflection symmetry $(AB)(CD)$ around the vertical axis,
  a variable symmetry in the model of Fig.~\ref{fig:queens} which can be eliminated here by constraining the first variable to be less than the last;
\item the horizontal axis reflection $(AD)(BC)$, i.e.~a value symmetry that can be eliminated here by constraining the first variable to be less or equal to the mid row value;
\item the diagonal reflection $(BD)$, a variable-value symmetry that can be broken by a lexicographic ordering constraint between the variables
  of the primal model and the dual model defined by $D_i=j\Leftrightarrow Q_j=i$;
\item the second diagonal reflection $(AC)$, similarly broken using a dual model;
\item the rotation by 90° $(ABCD)$ broken in the same manner;
\item the rotation by 180° $(AC)(BD)$;
\item the rotation by 270° $(ADCB)$;
\item (the identity).
\end{itemize}
These symmetry breaking constraints can be added to the model using lexicographic ordering \verb|lex_leq/2| constraints on arrays as follows:
\begin{verbatim}
sym_elim(N, Queens) :-
  Queens[1] #< Queens[N], % vertical reflection symmetry
  Queens[1] #=< (N+1)//2, % horizontal reflection symmetry
  
  int_array(Dual, [N], 1..N), 
  for_all([I, J] in 1..N, Queens[I] #= J #<==> Dual[J] #= I),
  lex_leq(Queens, Dual),  % first diagonal reflection
  
  int_array(SecondDiagonal, [N], 1..N),
  for_all(I in 1..N, SecondDiagonal[I] #= N + 1 - Dual[N+1-I]),
  lex_leq(Queens, SecondDiagonal), 

  int_array(R90, [N], 1..N),
  for_all(I in 1..N, R90[I] #= Dual[N+1-I]),
  lex_leq(Queens, R90),   % rotation symmetry by 90°

  int_array(R180, [N], 1..N),
  for_all(I in 1..N, R180[I] #= N + 1 - Queens[N+1-I]),
  lex_leq(Queens, R180),
  
  int_array(R270, [N], 1..N),
  for_all(I in 1..N, R270[I] #= N + 1 - Dual[I]),
  lex_leq(Queens, R270).
\end{verbatim}

Implementing arrays in ISO-Prolog 
is straightforward using term meta-predicates \verb|functor/3| and \verb|arg/3|,
and reading/writing cells by unification
(backtrackable and non-backtrackable assignments using \verb|setarg/3| or \verb|nb_setarg/3| are also possible
but not considered here). 

On the other hand, implementing bounded quantifiers and iteration meta-predicates like \verb|for_all/2| above raises some issues
with the existing ISO-Prolog meta-predicates 
which have been defined for the provability semantics of Prolog,
before the advent of the answer constraint semantics of Constraint Logic Programming~\cite{JL87popl}.
This justifies a new implementation of control meta-predicates compatible with constraint solving libraries in Prolog,
and probably the definition of a second level of normalization for ISO-Prolog with attributed variables for constraint handling.
These aspects are discussed in Section~\ref{meta}.

\subsection{Performance Figures}
\begin{table}
      \caption{CPU time in seconds and reported number of logical inferences in SWI-Prolog for solving various variants of the N-queens problem,
    compared between the recursive program on lists and the mathematical model using subscripted variables and quantifiers.
  The last benchmark simply shows the logarithmic versus linear access times in large datasets represented by arrays versus lists.}\label{table}
  \begin{tabular}{|c|c|c|}
    \hline
    Benchmark &  CPU time in seconds &  Nb logical inferences  \\
    \hline
    \textbf{100-queens first solution}&&\\
    math model& 0.373 & 4,552,518 \\
    list program& 0.298 & 4,472,406 \\
    all-distinct math model& 3.611 & 63,565,626 \\
    all-distinct list program& 3.470 & 63,502,308 \\
    \hline
    \textbf{8-queens all 92 solutions}&&\\
    math model& 0.041 & 757,835 \\
    list program&  0.042 & 757,303 \\
    all-distinct math model& 0.297 & 4,025,643 \\
    all-distinct list program&  0.289 & 4,012,622 \\
    \hline
    \textbf{8-queens all 12 non-symmetrical sol.}&&\\
    math model& 0.045 & 667,850 \\
    list program& 0.048 & 664,708 \\
    all-distinct math model& 0.125 & 1,471,542 \\
    all-distinct list program& 0.128 & 1,509,989 \\
    \hline
    $10^2$ accesses in an array of size $10^2$ & 0.000 & 301 \\
    in a list of size $10^2$ & 0.000 &  544 \\
    $10^3$ accesses in an array of size $10^3$ & 0.001 & 3,001 \\
    in a list of size $10^3$ & 0.003  & 4,144 \\
    $10^4$ accesses in an array of size $10^4$ & 0.005 & 30,001 \\
    in a list of size $10^4$ & 0.161 & 40,001 \\
    $10^5$ accesses in an array of size $10^5$ & 0.027 & 300,001 \\
    in a list of size $10^5$ & 13.615  & 400,001 \\
    $10^6$ accesses in an array of size $10^6$ & 0.148  & 3,000,001 \\
    in a list of size $10^6$ &  1567.177 & 4,000,144 \\
    \hline
      \end{tabular}
\end{table}

We report here some performance figures obtained with our modeling library in SWI-Prolog version 9.0.4,
including the \verb|clpfd| library of constraints over integer variables,
on a MacBook Pro 2,3 GHz Quad-Core Intel Core i7 32 GB 3733 MHz.

Table~\ref{table} shows the absence of significant difference in computation time between the execution of
our  models based on arrays and universal quantifiers for iterations,
and the Prolog programs based on lists and list recursions.

On the other hand, the speed-up due to logarithmic access in an array versus linear access in a list
(which does not occur in the list based model of the N-queens problem) begins to show up from size 1000 in pure cell access problems.

\section{Comparison to MiniZinc Modeling Language}\label{zinc}

\subsection{Types}

Compared to MiniZinc, the writing of the N-queens model is very similar:
\begin{verbatim}
int: n = 8;
array [1..n] of var 1..n: queens;
constraint forall (i in 1..n-1, d in 1..n-i)
                  (queens[i] != queens[i+d] /\
                   queens[i] != queens[i+d] + d /\
                   queens[i] != queens[i+d] - d);
solve satisfy;
\end{verbatim}
Type declarations are necessary in MiniZinc to overload operators and constraint predicates
over the four constraint domains considered in MiniZinc: the integers, the Booleans, the real numbers (floating points)
and the domain of finite sets of integers.

This is not implemented in our modeling library in Prolog since constraint predicate symbols are currently not overloaded.
More specifically
\begin{itemize}
\item
  constraint predicates over integers and Booleans values 0/1 of library \verb|clpfd| are prefixed and distinguished by the \verb|#| symbol,
\item
  constraints over real numbers are distinguished by enclosing between curly brackets in the \verb|clpr| library used,
\item
  constraints on finite sets are currently not implemented.
\end{itemize}

It is worth noting however that prescriptive typing and type inference are possible in Prolog,
as demonstrated for instance in~\cite{FC01TPLP}
using a powerful system of subtyping constraints~\cite{CF02iclp}.
This could be used to parse MiniZinc models and type check mixed MiniZinc-Prolog programs.

\subsection{Programming Search}

The search procedure is the second most important component of constraint programming techniques to solve hard combinatorial problems.
For that reason, MiniZinc offers the possibility to provide a limited set of search annotations,
which can be interpreted by the solvers,
for specifying and-choice heuristics (e.g.~first-fail principle), or-choice heuristics (e.g.~best-first criterions)
and also some search strategies such as dichotomic search.

Of course, such search options cannot cover all needs for controlling search.
In~\cite{MFS15ppdp}, it is however shown how some more elaborated search strategies such as Limited Discrepancy Search~\cite{HG95ijcai}
can be specified by constraints in the model.
In~\cite{RGST15cp}, a solver-independent language is proposed to control search in MiniZinc
with a limited set of instructions to execute each time a solution is found.
This is illustrated by the implementation of Large Neighborhood Search (LNS), an important strategy for optimization problems.

In~\cite{STWSS13constraints}, search combinators are proposed to specify general search instructions to execute at each choice point in the model.
This requires interaction with the solver at each node of the search tree
which is hardly supported by constraint solvers implemented in a different programming language.

The capability to program search at that level of granularity
can however make a decisive difference in hard combinatorial problems. For instance, in disjunctive scheduling,
it can be worth spending time to determine the best mutual exclusion constraint to select for the next choice point,
in order to prune the search tree by constraint propagation, instead of just duplicating the real search space with a bad selection.
Sophisticated strategies, including looking-ahead techniques for determining that most informative disjunctive constraint by trying them all
may thus become necessary to implement.

This requires the recourse to a programming language which is possible with a programming library-based approach to modeling,
like here in Prolog, a natural choice for relation-based modeling,
and not possible with a pure modeling language approach like MiniZinc or other algebraic modeling languages,
beyond a limited set of search annotations.

\subsection{Answer Constraint Semantics in Fourier's Example}

Thanks to the computed answer constraint semantics of the class of constraint logic programming languages~\cite{JL87popl},
and of Prolog with its constraint-solving libraries,
the execution of MiniZinc models directly in our modeling library in Prolog
makes it possible to obtain much more informative answer constraints, compared to the ground solutions obtained with MiniZinc solvers.

\begin{figure}
  \centering\includegraphics[width=0.5\textwidth]{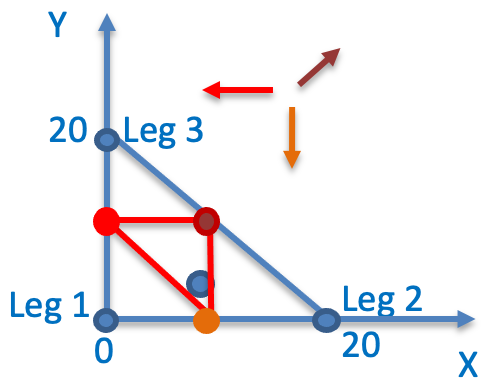}\caption{Example given by Fourier to illustrate a general purpose constraint-based modeling method,
  applied to the placement of a weight on a triangular table with limit constraints on the forces exerted on its legs 1, 2, 3,
  together with a general purpose method for solving systems of linear inequalities over the real numbers.
  The inner triangle represents the validity domain of the placement of a weight of 2 units.
This triangle is symbolically represented by the computed answer constraint for the query \texttt{fourier(2, X, Y, 1)}.}\label{fourier}
\end{figure}

This can be illustrated by the example given by Fourier in his lecture on constraint-based modeling
and solving for systems of linear inequalities~\cite{LM92jar}.
He took the problem of placing a given weight $P$ at unknown coordinates $(X, Y)$
on an isocele triangular table with a maximum force $F$ exerted on each of the 3 legs (see Fig.~\ref{fourier}).

By assuming that the triangle table has leg 1 at coordinates $(0, 0)$, leg 2 at coordinates $(20,0)$ and leg 3 at coordinates $(0,20)$,
simple moment equilibrium equations give the following model:

\begin{verbatim}
fourier(P, X, Y, F):-
    float_array(Forces, [3], 0..F),
    {Force[1]+Force[2]+Force[3] = P},
    {P*X = 20*Force[2]},
    {P*Y = 20*Force[3]}.

?- fourier(3, X, Y, 1).
X = Y, Y = 6.666666666666667.

?- fourier(3.1, X, Y, 1).
false.

?- fourier(2, X, Y, 1).
{Y=20.0-10.0*_A-10.0*_B, X=10.0*_B, _=2.0-_A-_B, _B=<1.0, _A=<1.0}.

?- fourier(2, X, Y, 1), maximize(X).
X = 10.0,
{Y=10.0-10.0*_A, _=1.0-_A, _A=<1.0, _A>=0.0}.

?- fourier(2, X, Y, 1), maximize(X+Y).
X = Y, Y = 10.0.
\end{verbatim}

For a weight of 3 units, a single point is returned as unique solution.
Interestingly, for the placement of a weight of 2 units maximizing its coordinate along the X axis,
the value 10 is found for X together with one constraint on Y and A which defines the vertical segment where the weight can be placed.
Similarly, the computed answer constraints returned for the placement of a weight of 2 units without optimization criterion
define symbolically the validity domain depicted by the inner triangle in Fig.~\ref{fourier}.
It is worth noting that the computed answer constraint set semantics are a quite unique advantage of the constraint logic programming paradigm,
compared to other declarative modeling languages that are usually interfaced to solvers restricted to compute ground solutions
and not constraints.

\subsection{Model Debuging and Visualization in a Unique IDE}

In our modeling library approach in Prolog, a MiniZinc model can be executed in two ways:
\begin{itemize}
  \item
either by parsing the MiniZinc syntax and interpreting it directly by the predicates of our library,
\item
  or by parsing the FlatZinc model generated by the MiniZinc compiler
which is the standard way of evaluating backend MiniZinc constraint solvers.
\end{itemize}

A FlatZinc parser has been developed for SICStus Prolog which shows excellent performance on MiniZinc challenge competitions~\cite{zinc}.

On the other hand, the advantage of parsing MiniZinc syntax in a Prolog system
has already been demonstrated in a system like Eclipse~\cite{AW06book}.
Making it available in a library for standard Prolog systems is however a significant contribution
to factorize system development efforts, and use the single Integrated Development Environment (IDE) of Prolog to debug MiniZinc models.

\begin{figure}
  \centering\includegraphics[width=0.6\textwidth]{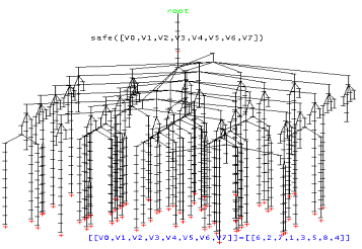}
\caption{Visualization of the search tree for enumerating all the 92 solutions of the 8-queens problem in Prolog with constraints,
  using the interactive graphical visualization system CLPGUI of~\cite{FSC04constraints}.
  This illustrates the good job of constraints
  for pruning the search tree in this example.
  Indeed, after around 5 choice points with a limited number of possible values,
  the branches of the search tree become linear and lead to one solution.}\label{8queens}
\end{figure}

For instance, the general purpose search tree visualization and interaction library CLPGUI, developed for constraint logic programming~\cite{FSC04constraints},
can be used here to visualize, and restore states in, the search tree developed by our modeling library in Prolog.
This is illustrated in Fig.~\ref{8queens} in the model above for enumerating all 92 solutions of the 8 queens problem.

From the point of view of a modeler, or of a teacher, this is a very useful tool that
is difficult to implement without a unique declarative programming/modeling IDE.
For the same reasons, model debuging is greatly facilitated by the absence of a first transformation of the model to a FlatZinc model,
followed by a second transformation to an external constraint solver which can create runtime errors difficult to recover.

\section{Modeling Library}\label{lib}

\subsection{Subscripted Variables with Library \texttt{arrays.pl}}

Subscripted variables, i.e.~arrays, are not normalized in ISO-Prolog but can be easily represented by terms with functor \texttt{array}.
Multidimensional arrays are then represented by arrays of arrays.
On top of that, library \texttt{modeling.pl} defines predicates for defining arrays of Booleans, integers or real number with domain constraints.
The main predicates defined in library \texttt{arrays.pl} are:

\verb|array(?Array, ?Dimensions)|
true if Array is an array of dimension Dimensions.
Either creates an array of given dimensions, or returns the list of dimensions of a given array, or checks the relation.

\verb|cell(+Array, +Indices, ?Cell).| true if Cell is the cell or subarray cell of a given array at given indices (integer or list of integers).
Throws an error if the indices are out of range of the array.
The cell is read and written by unification.
The functional notation syntax \verb|Array[Indices]| is also authorized in constraints to denote the cell of an array without
having to introduce an existentially quantified variable for that, e.g.~with the quantifier meta-predicates described below.

\verb|array_list(?Array, ?List).| true if List is the flat list of all elements in Array.
Either creates a one-dimensional array, or the list of array cells, or checks the consistency of both representation.

\verb|array_lists(?Array, ?List).| true if List is the list of lists of elements in the array following its dimensions.

\verb|tensor(+A, +Op, +B, +Rel, ?C)| equivalent to (A Op B) Rel C,
where A, B, C are arrays or lists of same dimensions, Op is a binary operation executed element-wise, and Rel a binary predicate.

Furthermore, for the sake of generality outside the scope of our modeling library,
backtrackable and non-backtrackable imperative array cell assignments are also defined
by extra-logical predicates \verb|set_cell/3| and \verb|nb_set_cell/3| using ISO-Prolog predicate \verb|setarg/3|.

\subsection{Bounded Quantification Meta-Predicates in \texttt{quantifiers.pl}}

Iteration with subscripted variables is more natural to implement using quantifiers on the indices rather than recursion.
To this end, we introduce the following general-purpose meta-predicates to quantify variables either universally or existentially,
and distinguish them from context variables in goals:

\verb|for_all(+Args, +Goal)| calls Goal for all the arguments specified in Args, i.e.~calls the conjunction \texttt{Goal(Arg1), ..., Goal(ArgN)}.
The list of arguments Args is a list of universally quantified variables given with a finite domain 
(either finite interval of integers or list of terms), plus possibly a deterministic condition goal, using expressions of the form
%
\verb|Args in Domain where Condition|.

\verb|exists(+Vars, +Goal)| to make a list of variables renamed-apart and local to the goal.

\verb|let(+Bindings, +Goal)|
to additionally apply a binding constraint \texttt{=, =.., in, \#=, \#<, \#>, \#=<, \#>=}
between an existentially quantified variable and a term, possibly using \texttt{Array[Indices]} functional notation for array cells.

\verb|list_of(+Vars, ?List)| to build the list of terms specified in Args.

The introduction of those meta-predicates in our modeling library is justified by the fact
that ISO-Prolog meta-predicates  \verb|bagof/3|, \verb|setof/3|, \verb|findall/3|
have been introduced before the advent of constraint logic programming, and of the implementation of constraint solvers based on attributed variables,
at a time focussing on the provability semantics of Prolog.

Indeed, in his seminal paper~\cite{Warren82mi}, David Warren introduced an extra logical meta-predicate to collect information
across branches of the search tree. Paraphrasing him, the extension takes the form of a new built-in predicate:
\verb|setof(X,P,S)|
to be read as:
\emph{"The set of instances of X such that P is provable is S"}.
This is achieved with a special mechanism introduced in the WAM~\cite{Warren83sri} for copying and memorizing terms in a list across backtracking.
The non deterministic goal P can thus be used as a \emph{generator of the instances} to be checked for satisfiability.

This mechanism has been used to introduce various meta-predicates 
for higher-order programming in Prolog~\cite{SW95iclp}.
Those meta-predicates thus refer to the provability semantics of Prolog, not the answer constraint semantics.
Similarly, meta-predicates \verb|forall/3|, \verb|foreach/3| found in several Prolog dialects
have no normalized behaviour with respect to attributed variables and constraints.

In the success set semantics of a universal quantifier for goals,
it is not clear whether the successfull bindings should be kept across backtracking.
On the other hand, in the answer constraint semantics, it is clear that the computed answer constraints should be added conjunctively to the store of constraints.
This is precisely what is achieved by our \verb|for_all/2| meta-predicate that generates instances of the goal by iteration,
unlike for example the \verb|forall/2| predicate of SWI-Prolog that generates instances by backtracking:
\begin{verbatim}
?- L=[X, Y], forall(member(V, L), V=a).
L = [X, Y].

?- L=[X, Y], for_all(V in L, V=a).
L = [a, a],
X = Y, Y = a.
\end{verbatim}

\subsection{Interface to Constraint Solvers with Library \texttt{clp.pl}}\label{meta}

We also found it useful to create a front-end interface to existing libraries for solving constraints over the real numbers, integers and booleans,
for several reasons.
First, this front-end library makes it possible to use evaluable expressions and \texttt{Array[Indices]} functional notation
in constraints and in the definitions of variable domains.
Second, global constraints on lists of variables can now undifferently accept arrays instead of lists.

Beyond that, some questions arise concerning
the answer constraint semantics and the copying or not by Prolog meta-predicates
of the constraints attached to the variables.
For constraint libraries using attributed variables, this amounts to the question of whether those meta-predicates should copy the
attributes or not, in particular the constraint propagator attributes.

In the answer constraint semantics, the answer is clearly yes, and this is what is done for instance in SWI-Prolog and Scryer-Prolog.
There is however currently a side effect concerning the duplication of the constraint propagators, e.g.

\begin{verbatim}
?- L=[A, B], L ins 1..2, A #=< B, bagof(W, member(W, L), L2).
L = L2, L2 = [A, B],
A in 1..2,
B#>=A,
B#>=A,
B#>=A,
B in 1..2.
\end{verbatim}

This is patched in our library \texttt{clp.pl} by making a set union instead of concatenating the constraint propagators attached to two variables when they are unified.
In this respect, it is worth noting that the set union of constraint propagators should not be implemented by maintaining the constraint propagation attributes sorted, since that could severely affect
the scheduling and performance of constraint propagators.

Furthermore, a more efficient and more general solution would be to consider the simplification of constraints on two variables that get unified.
This is not usually done in constraint programming and global constraint solvers, e.g.
\begin{verbatim}
?- L=[X, Y], L ins 1..3, all_distinct(L), X=Y.
L = [Y, Y],
X = Y,
Y in 1..3,
all_distinct([Y, Y]).
\end{verbatim}
Symbolic simplification of constraints is however instrumental in SMT solvers or in Constraint Handling Rules (CHR)~\cite{Fruhwirth09book}.
This has been shown responsible for drastic performance improvement in some contexts,
for instance in~\cite{CF02iclp,CF03fsttcs} for solving subtyping constraints using CHR, with better performance than dedicated algorithms in CAML,
thanks to the constraint simplifications performed by CHR immediately upon unification of two variables.

\subsection{Proposal for a Second-level of Normalization of ISO-Prolog}

Because of the importance of constraint-based methods in many applications of Prolog,
and because of the numerous implementations of constraint solvers in Prolog libraries using attributed variables,
we propose to specify the behaviour of ISO-Prolog predicates with respect to attributed variables
in a new level 2 norm for ISO-Prolog.

More specifically, we mainly propose to open discussion to:
\begin{enumerate}
\item
  normalize attributed variables;
  \item
specify the copying or not of attributed variables in ISO-Prolog predicates
\verb|findall/3, bagof/3, setof/3|;
\item define meta-predicates compatible with constraint answer semantics;
\item normalize predicates for arrays.
  \end{enumerate}

\section{Conclusion}

Last year was the 50th year of Prolog, and this remarkable longevity for a programming language
was clearly recognized as a mark of the fundamental importance of the relational programming paradigm based on
first-order Horn clause logic and constraints in decidable theories.
It also showed the need to transmit knowledge to new generations of Prolog system developpers, teachers and users,
and unite the Prolog community with new momentum.

The development of our constraint-based mathematical modeling library in Prolog aims at going in that direction
from the triple points of view of the users, by providing a higher-level modeling library with full programming features;
of the teachers, by providing a unique environment for both constraint-based modeling and programming;
and of the system developpers, by factorizing library development efforts,
and providing use cases of standard predicates and libraries that require some corrections
and probably a normalization effort.

On-going work concerns the addition of a MiniZinc parser to our modeling library in order to directly execute and debug MiniZinc models
in a single modeling/programming environment,
compare performances using the large database of MiniZinc models,
compute answer constraints not just ground solutions for getting more declarative answers,
and stop being blocked from programming search for hard decision making problems.

\subsubsection*{Acknowledgments.}
I am grateful to my students at Ecole Polytechnique for their interest in my courses on Constraint Programming
  including practical work that evolved over the years from Prolog to MiniZinc and now back to Prolog;
  to Mathieu Hemery and Sylvain Soliman for their participation in the teaching and fruitful discussions;
 to Guy-Alain Narboni for his vision of the importance of the Prolog heritage and the organization of the 50th year of Prolog in Paris;
 to Markus Triska, Ulrich Neumerkel and Christian Jendreiko for their organization of the Scryer Meetup in Dusseldorf;
 and to the reviewers for their comments.

%
%
%
\bibliographystyle{splncs04}
\bibliography{contraintes}

\end{document}